\documentclass[aps,pra,twocolumn,groupedaddress]{revtex4}
\usepackage{graphicx}
\usepackage{hyperref}
\begin{document}
\title{Application of an all-solid-state,
frequency-doubled Nd$:$YAP laser\\ to the generation of twin beams
at 1080 nm}
\author{Ruixiang Guo}
\author{Julien Laurat}
\author{Jiangrui Gao}
\author{Changde Xie} \author{Kunchi Peng} \affiliation{State Key
Laboratory of Quantum Optics and Quantum Optical Devices,
Institute of Opto-Electronics, Shanxi University, Taiyuan, Shanxi,
China 030006}

\begin{abstract}
A laser-diode-pumped intracavity frequency-doubled Nd$:$YAP KTP
laser is presented. Over 110 mw of TEM00 single-frequency output
power at 540-nm wavelength was obtained. The output green laser
was employed to pump a semimonolithic nondegenerate optical
parametric oscillator to produce intensity quantum correlated twin
beams at 1080 nm, and the maximum quantum noise squeezing of 74
$\%$(5.9 dB) on the intensity difference fluctuation between the
twin beams is observed. The threshold was reduced and the
stability was increased significantly when compared with similar
lamp-pumped systems.
\end{abstract}
\maketitle

\section{Introduction}
Nonclassical state light with strong quantum correlation is an
important light source for quantum measurements and quantum
information process. Continuous wave optical parametric
oscillators (OPOs) are among the most efficient devices likely to
produce nonclassical states of light such as the generation of
bright twin beams and quadraturesqueezed state of light
\cite{1,2,3,4,5}. The nonclassical state of light generated by OPO
has been successfully utilized in sub-shot-noise measurement,
quantum nondemolition measurement, quantum teleportation, and
quantum secret communication \cite{3,6,7,8}. However, aside from
the improvement of the degree of noise suppression, a central
issue is the establishment of nonclassical light generation
devices with long-term stability, compactness, and high efficiency
for practical application. With the arrival of high-power output
diodes, the prospect of an all-solid-state source as an
alternative to the arc-pumped laser has shown its special
advantage \cite{9}. The commercial availability of diode-pumped
monolithic miniature Nd$:$YAG lasers \cite{10} with their high
intrinsic frequency stability has opened up new opportunities for
squeezed-light generation at 1064 nm and its harmonics
\cite{11,12,13}. However, it is impossible to realize frequency
doubling or degenerate frequency downconversion in KTP crystals
through type-II noncritical phase matching using a Nd:YAG laser as
a pump source\cite{8}. Following the report of Garmash et al.
\cite{14} about type II noncritical phase matching in an
$\alpha$-cut KTP crystal at 1080 nm, several groups designed the
arc-lamp-pumped frequency-doubled and stabilized Nd:YAP (Nd:
YAlO3) laser operating at 1080 and 540 nm and accomplished
excellent experiments on nonclassical light generation and
application \cite{1,2,15,16}. Although several groups have worked
on the laser-diodepumped Nd$:$YAP laser with fundamental wave
output at 1080 nm \cite{17,18,19,20} a diode-pumped intracavity
frequency doubling YAP$/$KTP laser with singlefrequency output of
green light at 540 nm has not been published to our knowledge.
Indeed, the design of the intracavity frequency-doubled laser is
much more complicated. Here we present a compact Nd:YAP$/$KTP
laser, longitudinally pumped by a single 2.5-W cw diode bar. Over
110 mW of singlefrequency output power at 540 nm wavelength with a
threshold pump power of 360 mW and a slop efficiency of
approximately 8$\%$ were obtained. The fluctuation of power is
less than $\pm$1$\%$ during three hours, and the frequency float
is less than 3 MHz$/$min at the case of free running. The laser
was employed to pump a single semimonolithic cw nondegenerate OPO,
consisting of an $\alpha$-cut KTP crystal for type-II noncritical
phase matching at 1080 nm and an concave output coupler with the
transmission $T_1( 1080 nm)= 3\%$ and $T_2(540 nm)=11\%$. The
oscillation threshold of NOPO was only 3.7 mW, and a maximum
intensity-difference squeezing of 5.9 dB (74$\%$) are observed at
3 MHz; if the efficiency of the detection system is taken into
account, the actual squeezing of output light from the NOPO
\cite{4} should be 81$\%$ (7.2 dB). Compared with our lamp-pumped
system of twin-beam generation \cite{2}, the threshold was
significantly reduced, the quantum correlation between twin beams
was increased, and the stability was improved. The all-solid-state
system can operate stably to generate nonclassical light over one
hour when the active frequency-stability system of the OPO is
turned on \cite{2}.

\section{Experimental arrangements and results}
\subsection{All-Solid-State Frequency-Doubled Nd:YAP$/$KTP Laser}

\begin{figure}[htpb!]
\centerline{\includegraphics[width=.8\columnwidth]{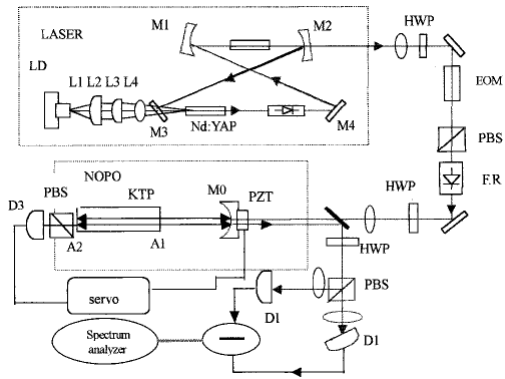}}
\caption{Experimental scheme of an all-solid-state
frequencydoubled Nd:YAP laser and NOPO for generation of twin
beams at 1080 nm. Both the laser and the NOPO are laid in a sealed
box with clean air. EOM, electro-optic modulator; PZT,
piezoelectric transducers; F.R, Faraday rotator; HWP, half
wavelength plate; P1, P2, P3, polarized-beam-splitter; D1, D2, D3,
detectors.}
\end{figure}

The experimental setup for all-solid-state Nd:YAP$/$KTP laser and
the generation system of twin beams at 1080 nm are shown
schematically in Fig. 1. The base of the laser housing is an invar
steel structure on which a four-mirror (M1–M4) bow-tie ring cavity
was built. The laser cavity consists of two concave mirrors with
the radius of curvature $R=50mm$ (M1,M2) and two plane mirrors
(M3,M4). All mirrors have a reflectivity of more than 99.5$\%$ at
1080 nm. M1 and M2 are separated by 57 mm, and M3 and M4 are
separated by 120 mm; the total length of the ring cavity is 330
mm. The incident angles on M1 andM2 are made as small as possible
to reduce both the astigmatic from the concave mirrors and the
polarization imbalance of the KTP crystal. The M3 and M2 are the
input and output couplers with transmissions of 90$\%$ for the
pump light at 802 nm and of 90$\%$ for the second harmonic wave at
540 nm, respectively. This geometry gives a $1/e^2$ beam radius of
approximately 40 $µm$ between the two concave mirrors where an
$\alpha$-cut KTP frequency doubler of $3x3x10 mm^3$ is located and
the KTP is heated to the phasematching temperature of about
63$^{\circ}$C with a precision of $\pm0.01^{\circ}$C. A Faraday
rotator and a halfwavelength plate ($\lambda/2$) for 1080 nm is
placed in the cavity to enforce unidirectional operation. The main
feature of laser is a b-axis cut Nd:YAP rod that is 2 mm and 4 mm
long and pumped by a 2.5-W diode-laser bar (Model
S-81-2700c-200-h, Coherence Corporation, 80 Rose Orchard Way, San
Jose, Calif. 95134) with a coherent radiation of about 805 nm at
25 °C from a 200 $µm$ x 1 $µm$ junction. In the parallel and the
perpendicular directions to the junction, the beam divergences are
$\theta_\parallel= 7.45^{circ}$ and $\theta_\perp= 2.06^{circ}$
for a FWHM, respectively. The light is collimated and focused onto
the YAP crystal by a self-focus rod (L1, pitch $= 0.22$), two
cylindrical lenses (L2, f $=$ 20 mm; L3, f $=$ 40 mm) and one
spherical lens (L4, f $=$ 50 mm). After the transmission from the
beam collimated system, the pump power before M3 is about 1.8 W.
The collimated farfield spot of the pump beam was nearly square
with the beam divergence of 1.2 mrad. The diode beam was focused
on the crystal with a spot size of approximately 60 $µm$ x 60
$µm$, and a divergence of 60 mrad. The 1$\%$-doped YAP crystal is
placed between the two plane mirrors, taking into account the
laser rod with a thermal focal length about 300 mm at 1.8 W pump
power, and the beam waist in YAP is about 200 $µm$ (half width),
so that all pump light is confined in the TEM00 of laser. The YAP
crystal is an optical biaxial crystal, thus both the polarized
absorption spectra and the polarized fluorescence spectra present
obviously anisotropic characteristics. The peak absorption and the
peak emission were found when the polarization of pump field was
parallel with the c-axis, where the peak absorption is 14
cm$^{-1}$ at 802 nm and the peak emission cross section is 3.0
$10^{-19} cm^2$ at 1080 nm \cite{15}. Adjusting the orientation of
YAP crystal to make its c axis parallel to the polarization of
pump light and tuning the temperature of the laser diode at 14.5
$^\circ$C to emit the wavelength of 802 nm, we obtained the
maximum pump absorption efficiency of more than 96$\%$. In the
design of the laser, great care was taken to reduce the
intracavity losses for higher conversion efficiency, and we
obtained some benefits from the anisotropic characteristics of the
YAP crystal. For example, no polarized component and frequency
choosing component are needed in the cavity, and therefore the
intracavity loss is decreased to the largest extent; the total
round-trip cavity loss owing to the insertion loss of
antireflection-coated components (half-wavelength plate, KTP, YAP
and TGG crystal) and the residual loss of the four high-reflection
cavity mirrors is estimated to be about $\sim 3\%$.

The threshold of laser oscillation was reached when the pump power
on the laser rod was about 360 mW, and a single-frequency of over
110 mW output power at 540-nm wavelength and a slop efficiency of
approximately 8$\%$ were obtained at the laser pump power of 1.8
W. The losses of the collimate optics and intracavity components
influence the improvement of efficiency. The fluctuation of power
is less than $\pm$1$\%$ during three hours, and the typical
scanning Fabry-Perot trace for the 1080-nm leakage (Fig. 2)
confirmed its single-frequency operation with a frequency float of
3 MHz$/$min at the case of free running, and the laser beam was
TEM00 with typical $M^2$ values of 1.14 and 1.03 in the horizontal
and vertical planes, respectively.

\begin{figure}[htpb!]
\centerline{\includegraphics[width=.8\columnwidth]{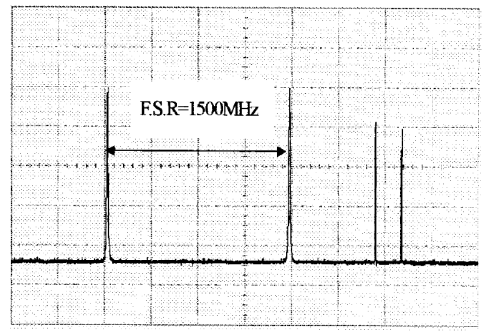}}
\caption{Typical scanning Fabry-Perot trace for the 1080-nm
leakage while the power of 110 mW is generated at 540 nm,
demonstrating the single-frequency operation of laser. F.S.R in
figure stands for free spectral range.}
\end{figure}

\subsection{Triply Resonant Nondegenerate OPO}
The triply resonant NOPO contains a 7-mm-long $\alpha$-cut KTP
crystal heated to approximately 63 $^\circ$C for type-II
noncritical phase matching to generate quantum correlated twin
beams at 1080 nm through a frequency downconversion process and a
concave mirror M0 with 30-mm radius of curvature that is used as
both an input and an output coupler. The KTP crystal is polished
flat and antireflection- and high-reflection-coated at both 540 nm
and 1080 nm on the facets A1 and A2, respectively. The concave
mirror with transmission $T_1 (1080 nm) = 3\%$ and $T_2 (540
nm)=11\%$ is placed 23 mm from the A1 facet of crystal. The
semimonolithic cavity is chosen for its mechanical stability and
slight intracavity loss. The escape efficiency of 0.893,
calculated from transmission T1, and the total dissipate loss of
0.36$\%$ for 1080 nm, which is measured from the finesse of
cavity, imply a theoretical squeezing of 7.8 dB at the analyzing
frequency of 3 MHz (see Eq. 1 in Ref. 4). The pump light at 540 nm
from the laser passes through an isolator (F.R) and is carefully
mode matched into the NOPO (the matching efficiency is better than
85$\%$). For our OPO with high finesse for the signal and idler
fields and low finesse for the pump field, there is at least one
pair of signal and idler modes that can resonate within a pump
resonance peak, therefore triply resonant OPO operation can be
fulfilled only through adjustment of the length of the cavity
\cite{21}. The cavity bandwidth is 22 MHz, which results in a
squeezing bandwidth from 0 MHz to 10 MHz without a significant
penalty in the noise reduction. The Pound-Drever-Hall method is
used to lock NOPO on resonance. The pump light is modulated
electrooptically with an electrooptic modulator at 19.2 MHz, and
the signal and the idler output from the facets A2 of the KTP
crystal pass through a half-wave plate and polarizer P3 polarizing
horizontally. Then the output is monitored by a high speed
detector D3. Since the electrooptical signal modulated on the pump
wave is transmitted to the generated signal and idler waves within
the cavity and can be detected on the leakage field leaving from
the facets A2 of the KTP crystal,22 mixing the detected signal
from D3 with the modulation source of the pump waves and low-pass
filtering then results in a dispersion-type error signal, that can
be fed back to the piezoelectric transducer mounted on M0 to lock
the cavity on-resonance by a servo loop (we use homemade
proportional-integraldifferential to provide the electronic
feedback control). The measured minimum oscillation threshold of
NOPO is about 3.7 mW in the case of exactly triple resonance. A
visible-to-infrared power conversion of 50$\%$ is demonstrated
under the pump power of 18.7 mW at 540 nm. The measured
wavelengths of the downconverted twin infrared beams at the
temperature of 63.54 °C are $\lambda_1 = 1080.030 nm$ and
$\lambda_2 = 1079.996 nm$. The signal and idler beams from M0 are
separated by a polarized beam splitter and then focused on a pair
of InGaAs pin photodiodes EpitaxxETX 500 $[$Epitaxx Inc., 7
Graphics Drive, West Trenton, N.J. 08628) D1, D2 diameter 500 $µm$
$]$ with a tilting of the Brewster angle to the polarization plane
of the monitored beams in order to reduce the losses of reflection
on the photodiode surface;15 their quantum efficiency is measured
to be 94$\pm$2$\%$. All of the optical surfaces between the OPO
cavity and the detectors are antireflection coated. The output
photocurrents from the detectors of the ETX500 diode is amplified
by a two-stage transimpedence amplification circuit with two
operational amplifiers (AH0013, Optical Electronics, Inc., 4455 S.
Park Avenue 106, Tucson, Ariz. 85714 and Mar6, Mini-Circuits,
Inc., P.O. Box 350166, Brooklyn, N.Y. 11235), and then subtraced
with a 180° power combiner (Minicircuit ZSCJ-2-1; the frequency
range is from 0.5 to 20 MHz). The noise on the resulting
difference is monitored by a spectrum analyzer. The linearity of
detection was verified by a dc saturation measurement, and this
photo detector can take up to 7.5mW of optical power without any
observable saturation effect. The measured electronic common mode
rejection ratio is 28 dB at 4 MHz. The measured electronic noise
floor of detection system is about 22 dB lower than the shot
noise. There is quite large excess noise (15 dB above shot noise
at 3 MHz) on signal and idler beams individually.

\begin{figure}[htpb!]
\centerline{\includegraphics[width=.8\columnwidth]{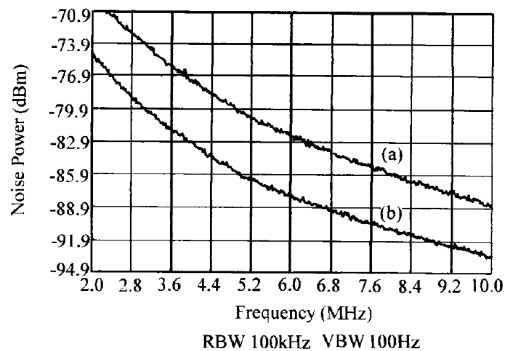}}
\caption{Noise power spectrum of intensity difference between twin
beams at a pump power of 18.7 mW. Resolution bandwidth, 100 kHz;
video bandwidth, 100 Hz. (a) shot-noise limit, (b) noise spectrum
of intensity difference between the twin beams.} \end{figure}

The experimentally measured squeezing spectrum of intensity
difference fluctuation between twin beams is displayed in Fig. 3,
in which trace (a) is the shot-noise limit measured through
rotation of the polarization of the output signal and idler beam
to 45° relative to the polarization direction of the polarized
beamsplitter4; trace (b) is the squeezed-noise power spectrum. The
measured maximum squeezing is 5.9 dB at 3 MHz and 7.2 dB after
taking into account the measurement efficiency of 90$\%$
(including the detector quantum efficiency of 94$\%$ and the
transmitting losses of 96$\%$), which is quite close to the
theoretically calculated value (7.8 dB).

\section{Conclusion}
 We designed and demonstrated a reliable and
compact all-solid-state single-frequency Nd:YAP$/$KTP laser for
the first time, and its output second harmonic wave was employed
as a pump source of a NOPO to produce the nonclassical light
field. Since the wavelengths of 1080 nm and its second-harmonic
generation 540 nm can perform the type-II noncritical phase
matching in an $\alpha$-cut KTP, the laser has important and
potential application in the nonclassical light generation and
future experimental quantum optical and quantum information. The
output transmission of 3$\%$ at 1080 nm used in experiments
limited the squeezing; if the transmission of coupler is increased
by e.g., 5$\%$, the larger squeezing over 10 dB will be
theoretically expected.

This work was supported by the National Nature Science Foundation
of China (approval 69837010) and the Younger Science Foundation of
the Shanxi Province (approval 20001015).

\end{document}